\documentclass[a4paper]{toledoStyle}
\usepackage{epsfig}
\begin{document}

\title{The Balloon-borne Large Aperture Sub-millimetre Telescope}

\author{Douglas Scott\inst{1}\\
and the BLAST Team:\\
Peter Ade\inst{2}, Jamie Bock\inst{3},
Paolo DeBernardis\inst{4}, Mark Devlin\inst{5}, Matt Griffin\inst{2},
Josh Gundersen\inst{6},\\
Mark Halpern\inst{1}, David Hughes\inst{7}, Jeff Klein\inst{5},
Silvia Masi\inst{4},
Phil Mauskopf\inst{\,8},
Barth Netterfield\inst{9},\\
Luca Olmi\inst{10}, Lyman Page\inst{11}, Greg Tucker\inst{12} } 

\institute{
  Dept.~of Physics \& Astronomy, University of British Columbia,
  6224 Agricultural Road, Vancouver, B.C.\ V6T1Z1, Canada
\and
 Queen Mary and Westfield College, London, UK
\and
 Jet Propulsion Laboratory, Pasadena, CA, USA
\and
 University of Rome, Rome, Italy
\and
 University of Pennsylvania, Philadelphia, PA, USA
\and
 University of Miami, Coral Gables, FL, USA
\and
 INAOE, Puebla, Mexico
\and
 Cardiff University, Cardiff, UK
\and
 University of Toronto, Toronto, ON, Canada
\and
 University of Massachusetts, Amherst, MA, USA
\and
 Princeton University, Princeton, NJ, USA
\and
 Brown University, Providence, RI, USA}
\maketitle 

\begin{abstract}

The Balloon-borne Large-Aperture Sub-millimetre Telescope (BLAST)
will operate on a Long
Duration Balloon platform with large format bolometer arrays at 250, 350 and
$500\,\mu$m, initially using a $2.0\,$m mirror, with plans to increase to
$2.5\,$m. BLAST is a collaboration between scientists in the USA, Canada, UK,
Italy and Mexico.  Funding has been approved and it is now in its building
phase.  The test flight is scheduled for 2002, with the first long duration
flight the following year.  The scientific goals are to learn about the
nature of distant extragalactic star forming galaxies and cold pre-stellar
sources by making deep maps both at high and low galactic latitudes.  BLAST
will be useful for planning {\sl Herschel\/} key projects which use SPIRE.

\end{abstract}

\section{Introduction}

The far-IR background was first discovered in the {\sl COBE\/} data 5
years ago (\cite{scottd:puget96}), and has since been estimated at several
wavelengths in the $100\,\mu$m to $1\,$mm range
(\cite{scottd:fixsen,scottd:hauser,scottd:lagache}).
As shown in Fig.~1, the Far-IR Background (FIB) represents the most significant
energy density of photons after the CMB, and is roughly a factor of two larger
than the optical/near-IR background (although this is still debated).
The FIB peaks around $100\,\mu$m, and appears to be wider than
the CMB (indicating perhaps that the sources come from a range of redshifts).
Unlike for the x-ray background, we only had to wait a year or two before
a significant fraction of the FIB had been resolved.

Several surveys with the SCUBA instrument on the James Clerk Maxwell
Telescope found a great many more sources than expected in no
evolution models
(e.g.~\cite{scottd:SIB,scottd:hughes,scottd:barger,scottd:lilly,scottd:chapman,scottd:borys}).
The most sensitive SCUBA
band at $850\,\mu$m favours high redshift dusty galaxies compared with those
providing most of the FIB.  Or in other words,
the background at these wavelengths is about a factor of 30
below the value at the peak.  Hence SCUBA does not tell us directly about
the bulk of the galaxies responsible for the FIB.
However, information about sources at the peak
itself has also recently came from surveys with ISOPHOT on the {\sl ISO\/}
satellite.
The FIRBACK survey, for example, has provided deep maps over several
square degrees at $170\,\mu$m which resolve about 10\% of the background
(\cite{scottd:puget99,scottd:dscott,scottd:dole}).

\begin{figure}[ht]
  \begin{center}
    \epsfig{file=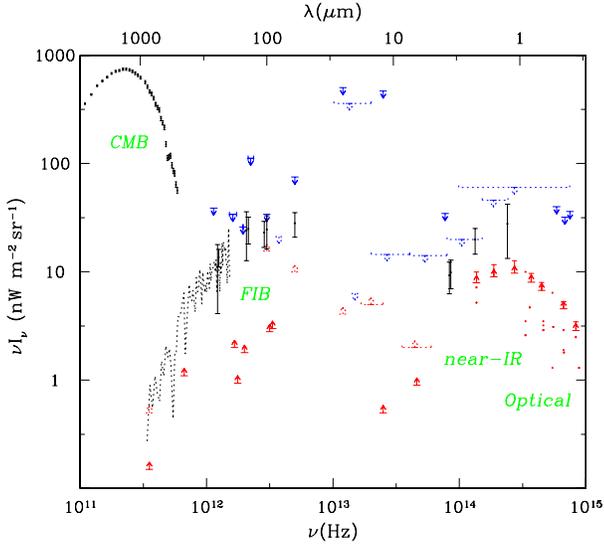, width=9cm}
  \end{center}
\caption{ Extragalactic background radiation measurements and constraints
from the cm-wave through to the optical region.  The infrared background
is beginning to be traced out over a wide wavelength range, and has a peak
near $100\,\mu$m. Here points with error bars represent measurements, while
arrows are upper or lower limits, and dashed lines show model dependent
limits.  Earlier plots with extensive references appear in Scott et al.~(2000b)
and Halpern \& Scott~(2000).}
\label{scottd_fig:fig1}
\end{figure}

Many questions remain however: What are the redshifts, star-formation rates
and morphologies of these FIRBACK galaxies?  What makes up the other 90\%
of the far-IR background?  How are these objects related to the SCUBA-bright
sources?  Are they the higher redshift equivalents of local luminous and
ultra-luminous infrared galaxies?  Is merging involved?  Are they related
to AGN activity?  In order to answer such questions we need to obtain
data at a wide range of different wavelengths, with the smallest possible
beam-sizes, to aid comparison between data sets.  A crucial part of this
puzzle is deep mapping at far-IR/sub-mm wavelengths which are near the
peak of the background.

This is one of the main motivations behind BLAST (Devlin et al.~2000).
At balloon altitudes
the atmosphere is essentially transparent at $250\,\mu$m (see Fig.~2), while
those wavelengths are impossible from the ground.
At $350\,\mu$m observations from
the best observatory sites are possible, but extremely challenging.
Working in the atmospheric window at $450\,\mu$m from sites like Mauna Kea
is considerably easier, but even there the atmosphere typically makes the
data 10 times noisier than at $850\,\mu$m.  BLAST will also have a channel at
$500\,\mu$m, which will be important for connecting to the ground-based
sub-mm observations.  But for BLAST,
pushing to the shorter wavelengths enables the beam-size to be smaller.
By flying a mirror of diameter $\ga2\,$m (significantly larger than
flown in most CMB balloon projects, for example), BLAST is able to
recover a beam-size which is similar to that of ground-based sub-mm
telescopes.

\begin{figure}[ht]
  \begin{center}
    \epsfig{file=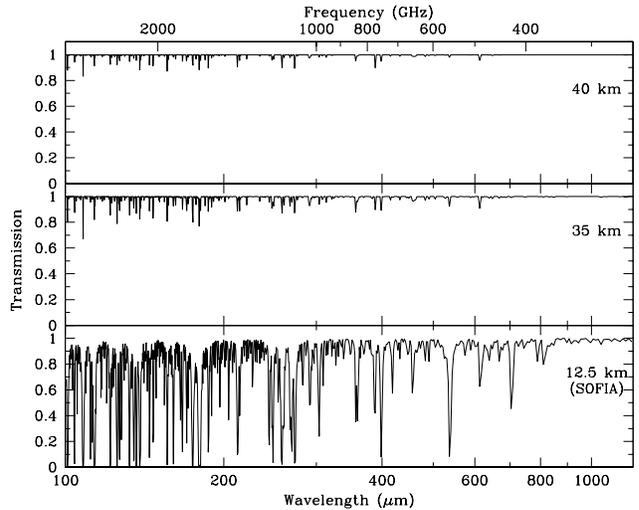, width=9cm}
  \end{center}
\caption{Atmospheric transmission at three different sub-orbital
altitudes.  At typical ballooning altitudes the atmosphere is essentially
transparent across this whole wavelength range.  This should be contrasted
with SOFIA altitudes and even the best high-altitude observatories (not
shown), where things are considerably worse.}
\label{scottd_fig:fig2}
\end{figure}

By carrying out deep surveys at high and low galactic latitudes, BLAST will
be able to address many related science goals, including:
\begin{enumerate}
  \item sub-mm continuum of solar system objects (planets, asteroids, \dots);
  \item study of cool dust in the ISM (pre-stellar objects, mass function
of clumps, \dots);
  \item nearby galaxies (distribution of cool dust, relation to star
formation, \dots);
  \item distant galaxies (number counts, resolving the far-IR background,
star formation history, \dots);
  \item correlations (clustering of far-IR galaxies, bias, merging, \dots);
\end{enumerate}

\section{Technical description}

BLAST is the Balloon-borne Large Aperture Sub-milli\-metre Telescope,
designed to map parts of the sky at sub-mm wavelengths which are essentially
impossible from the ground.  BLAST is a collaboration between 16 scientists
at a dozen institutions in 5 countries, with PI Mark Devlin at the
University of Pennsylvania.

\begin{figure}[ht]
  \begin{center}
    \epsfig{file=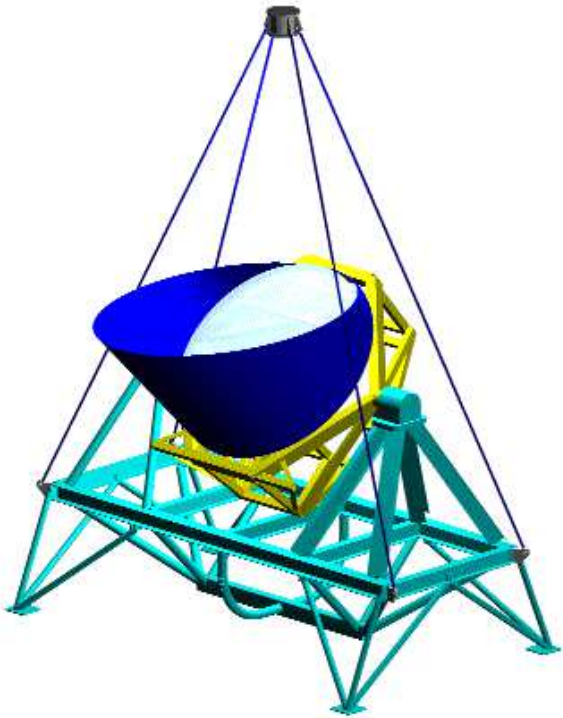, width=9cm}
  \end{center}
\caption{The BLAST telescope and balloon gondola design is shown here, with
sun shields and ground shields removed.  The alt-az pointing system will have
an absolute accuracy of better than $10^{\prime\prime}$, limited by
gyroscope and  daylight star-sensor performance.}
\label{scottd_fig:fig3}
\end{figure}

BLAST (see Fig.~3) will initially use an available $2\,$m diameter primary,
with plans
to upgrade to a larger  mirror for later flights.  Detectors will consist
of an array of 280 spider-web bolometers at three different wavelengths, with
most of the bolometers being at the shortest wavelength, and the entire field
of view being $6.5\times13$ arcminutes.  BLAST uses the same array structure
as the {\sl Herschel\/} SPIRE instrument.  The central
wavelengths will be $250\,\mu$m, $350\,\mu$m
and $500\,\mu$m, with beam-widths of about 30, 41 and 59 arcseconds,
respectively.
In a 6 hour map of 1 square degree BLAST will have a $1\sigma$ sensitivity of
about $15\,$mJy at each of the three wavelengths (see Table~1 for
detailed parameters).

\begin{table}[bht]
  \caption{Experimental parameters for BLAST.}
  \label{scottd:tab1}
  \begin{center}
  \leavevmode
    \footnotesize
    \begin{tabular}[h]{lccc}
      \hline \\[-5pt]
      Central wavelength & $250\,\mu$m & $350\,\mu$m  &  $500\,\mu$m \\[+5pt]
      \hline \\[-5pt]
      Number of pixels   & 149     &  88      &  43 \\
      Beam FWHM          & $30^{\prime\prime}$ & $41^{\prime\prime}$
                                                   & $59^{\prime\prime}$ \\
      NEFD (mJy.$\sqrt{\rm s}$) & 236 & 241 & 239 \\
      $\Delta S_\nu$[$1\sigma$,$1\,$hr,$1\,{\rm deg}^2$] (mJy) &38 &36 &36 \\
      \hline \\
      \end{tabular}
  \end{center}
\end{table}

Much of the emphasis for the BLAST team has been a fast delivery of Science,
and so the schedule calls for the first flight (from North America) in 2002,
with the first long-duration flight in 2003.

\section{BLAST Surveys}
\subsection{Deep cosmological surveys}
BLAST will carry out both Galactic and extra-galactic surveys.  For a
6 hour test flight we can map a region of $0.55\,{\rm deg}^2$ to the
confusion limit.  This
will be centred on one of the fields which have been well studied at
other wavelengths, e.g.~the Lockman Hole, HDF region or one of the ELAIS
ISO fields.  With rms sensitivity of about $10\,$mJy
at each wavelength, we might detect 150 sources at ${>}\,5\sigma$.
Combination of BLAST fluxes with those from other
instruments and facilities (e.g.~VLA, SCUBA, BOLOCAM, optical, {\sl CHANDRA},
{\sl XMM}) will allow the properties of the sub-mm luminous galaxies to be
studied across the full electromagnetic spectrum.  A long-duration flight
allows for a much larger region of tens of square degrees
to be mapped down to the confusion limit, and about 1500 galaxies
should be well detected.
Coordinating this with the {\sl SIRTF\/} SWIRE Legacy survey, for example,
would make scientific sense.

\subsection{Galactic Plane Surveys}

In the plane of the Milky Way BLAST will map a modestly-sized bright region
during the test flight.  A long-duration flight would allow for, say, a
$5^\circ\times10^\circ$ map of the Galactic Centre region, where there is
already extensive multi-wavelength data (e.g.Pierce-Price et al.~2000).
Combination of the BLAST map with data from SCUBA, {\sl MSX}, {\sl IRAS},
{\sl ISO}, the Canadian Galactic
Plane Survey, etc. would allow many different galactic science investigations:
\begin{itemize}
  \item studies of individual clouds and complexes;
  \item profiles and other properties of pre-stellar cores;
  \item mass functions of star-forming clumps;
  \item dus-to-gas ratio variations;
  \item studies of turbulence;
  \item correlations between $T_{\rm dust}$ and emissivity;
  \item grain properties.
\end{itemize}

\subsection{Statistics of Maps}
The traditional quantity derived from surveys is an estimate of the counts
of well-detected sources, $N({>}\,S)$.  This is the mainstay of
extragalactic studies in the far-IR/sub-mm, and a similar approach is used
in many Galactic studies also.  A great deal of theoretical modelling
of galaxy counts has been carried out
(e.g.~\cite{scottd:guid,scottd:pearson,scottd:franc,scottd:tan,scottd:tak,scottd:malkan}).
Detailed number counts can constrain the evolution of the various populations
of galaxies.  Here multi-wavelength studies (e.g.~BLAST ${+}$ SCUBA ${+}$
\dots) constrain the models much more effectively, and ultimately the models
are really pinned down when redshift data become available.

Number counts are clearly not the only statistic obtainable from a map.
This is particularly clear for dust maps at low galactic latitudes, where
there is structure on all scales, and the most physically motivated statistical
descriptors have yet to be firmly established.  The same is at least partly
true for extragalactic maps, but there it is expected that the bulk of the
information {\it will\/} be contained in one- and two-point statistics.
One point statistics, i.e.~looking at histograms of pixel intensities, allow an
estimate of source counts below the individual detection limit.  Usually
referred to as $P(D)$ analysis in radio and x-ray astronomy, this has been
applied to SCUBA data by Hughes et al.~(1998), for example.

Two-point correlation statistics have been used to look at faint sources
producing fluctuations in the far-IR background (\cite{scottd:lagache2000})
and in SCUBA maps (\cite{scottd:peacock}).  However, so far it is only the
Poisson fluctuations due to sources that are undetected individually that have
been measured -- correlations from clustering of the sources has proved
ellusive.  This is another direction in which BLAST could make substantial
progress.  The amplitude of clustering of sub-mm bright galaxies is
entirely unknown at this point.  Presumably they will be quite highly biased,
making them strongly clustered.  But if they come from a wide range of
redshifts their angular clustering will be partially  washed out.  Some
estimates have been made
(e.g.~\cite{scottd:scowhi,scottd:haikno,scottd:gazhug}), with the most thorough
theoretical investigation so far by Knox et al.~(2001).
These estimates indicate that clustering of the galaxies which comprise the
FIB can dominate over shot-noise at scales above about 10 arcminutes.
Measuring such correlations in the maps would provide additional constraints
on galaxy evolution models, bias, clustering, etc.
In general this is challenging, since it
requires making maps with no systematic effects at the largest scales.
However, this is certainly feasible for BLAST.

\section{Conclusions}
Clearly the SPIRE instrument on {\sl Herschel\/} will have extraordinary
capabilities for mapping at sub-millimetre wavelengths.  In the meantime,
however, it is possible to get more modest amounts of data much more cheaply
and much more quickly.  The data produced by BLAST should be very useful in
planning ambitious surveys with SPIRE.

\begin{acknowledgements}
DS would like to thank the organisers of the `Promise of {\sl FIRST\/}'
Symposium for a fun and intersting meeting.

\end{acknowledgements}

\end{document}